\documentclass[12pt,aps,pra,showpacs,floatfix]{revtex4}
\newcommand{\beq}{\begin{equation}}
\newcommand{\eeq}{\end{equation}}
\newcommand{\beqa}{\begin{eqnarray}}
\newcommand{\eeqa}{\end{eqnarray}}
\newcommand{\f}{\frac}

\newcommand{\kt}{\rangle}
\newcommand{\br}{\langle}
\newcommand{\ua}{\uparrow}
\newcommand{\da}{\downarrow}
\newcommand{\al}{\alpha}
\newcommand{\be}{\beta}
\newcommand{\ga}{\gamma}
\usepackage{graphicx}
\begin{document}
\title{What is random about a quantum random walk?}
\author{Arul Lakshminarayan}
\email[]{arul@prl.ernet.in}
\homepage[]{http://www.prl.res.in/~arul}
\affiliation{Physical Research Laboratory,
Navrangpura, Ahmedabad, 380009, India.}
\date{\today}
\begin{abstract} We use simple deterministic dynamical systems
as coins in studying quantum walks. These dynamical systems can be
chosen to display, in the classical limit, a range of behaviors from
the integrable to chaotic, or deterministically random. As an example
of an integrable coin we study the Fourier walk that generalizes the
Hadamard walk and show that the walker slows down with coin
dimensionality, which controls the effective Planck constant.
Introducing multi-Harper maps as deterministic models of random walks
we study the effect of coin chaos on the quantum walk. We also
demonstrate that breaking time-reversal symmetry in the coin dynamics
effectively slows down the walk.
\end{abstract}
\maketitle

\section{Introduction}

There have been many recent works on quantization of classical random
walks apparently with a view to implement them in quantum algorithms
analogous to the way classical random walks are used in classical
computation. For a recent excellent overview and references to the
literature we refer to
\cite{kempe}. While the simplest classical random walk has a
protagonist who decides whether to step to the right or left based
on the random result of a coin toss, head or tail, the quantum
counterpart's coin is a reversible unitary one, thereby making the
quantum walk apparently devoid of random elements. In the event of
evolution that is uninterrupted by measurement the quantum random
walk will then presumably be a quantization of a classical
deterministic system rather than a random one. Thus the classical
limit of these walks, which we may characterize as the
classical-coin limit may turn out to be deterministically random
ones, in other words those that display ``hard'' chaos. On the other
hand there has by and large been no conscious attempt at
incorporating this element in quantum random walks. Thus in this
paper we seek to understand the effect that classical
(non)integrability of the coin, which leads to its classical random
properties, has on the quantum walk. We will consider the
simplest case of a quantum random walk and study examples where
the coin dynamics ranges from regular to deterministic chaos in
the classical limit.

Consider a finite lattice ${|n \kt},$ where $n=0,\ldots,L-1$ and
$L$ is the number of sites. A particle hopping on this lattice is
endowed with an internal ``coin'' degree of freedom in a $M$
dimensional Hilbert space ${\mathcal H}_C^M$, apart from its state
on the lattice that belongs to a Hilbert space ${\mathcal
H}^{L}_P$. Let $P_R$ and $P_L$ be two projectors on the coin space
such that $P_R+P_L=I$. Defining a coin-flip by a unitary operator
on ${\mathcal H}^M_C$, say $U$, an example of a quantum random walk
is provided by the unitary operator on the product Hilbert space
${\mathcal H}^L_P \otimes {\mathcal H}^M_C$:
\beq \label{defineQRW}  E=\left(S \otimes P_R + S^{-1} \otimes P_L
\right) \left(I \otimes U \right). \eeq
Here $S$ shifts states on the lattice, $S|n \kt = |n+1\kt$. We
will in this paper consider periodic lattices, so that $S^L=I$.
The walk is given by iterating an initial state with the
above unitary operator and for instance the variance in the site
position could be monitored for diffusion.

Quantum random walks that have been studied for instance using the
``Hadamard'' coin lead to quadratic rate of diffusion, as opposed to
to the classical normal linear law \cite{Nayak}. Thus this is claimed to speed
up the walk \cite{kempe}, also the probability distributions are highly
oscillatory, and do not have an asymptotic limit, as opposed to
the limiting classical Gaussian distribution. Thus there are
considerable differences between a classical and quantum random
walk even for simple models and geometries.

We wish to consider using coins whose classical dynamics is understood
to be either random or regular. Deterministic randomness has of course
been largely studied as ``chaos'' and there are many models that are
decidedly deterministic but are rigorously isomorphic to random
processes such as Markov chains or Bernoulli processes
\cite{Ornstien}. An example is provided by the baker map \cite{LL}
that is a simple two-dimensional area preserving mapping which is as
random as a coin toss in the rigorous sense that it is isomorphic to
the $(1/2,1/2)$ Bernoulli process. Thus the quantum baker map
\cite{BalVoros} may be used as quantum coin dynamics in a random walk. It is
natural to consider quantizations of such systems as ideal models of
quantum random walks with well understood classical limits.

In fact the relevance of such systems to quantum random walks has been
already pointed out in the works of Wojcik and Dorfman
\cite{DanielDorfman,DD2}, where quantum multi-baker maps have been
studied. Multi-baker maps have in recent times been studied as exactly
solvable models of deterministic non-equilibrium transport processes
\cite{Gasp}. The kind of deterministic random walk models studied here
is a generalization of those that were first studied by Tasaki and
Gaspard
\cite{GT} who used the baker maps. Quantum multi-bakers
were first constructed and studied in \cite{ArulBal}, with a view
to understanding quantum effects on transport. The more recent
work \cite{DanielDorfman} adopts what was called a
``semiclassical'' approach in the earlier work, due to the
assumption of a product structure between ``site'' and ``internal''
degrees of freedom, which are really both in the same phase-space.
This results in neglecting a phenomenon akin to tunneling between
the sites even in the absence of an explicit shift. The work in
\cite{DanielDorfman} corresponds to a realization of the ``coined''
random walk as in Eq.~(\ref{defineQRW}), with $U$ being replaced by
the quantum baker map, with appropriately chosen projectors.  Although
this was not the explicit form in which the quantum multi-baker was
written therein, it corresponds to what the authors called a uniform
quantum multi-baker; there is only one coin that is used. We will in
this work choose the simpler ``semiclassical'' quantization procedure
as this leads to operators that are in the form of the simple
quantum walk in Eq.~(\ref{defineQRW}).

\begin{figure}
\includegraphics[height=2in]{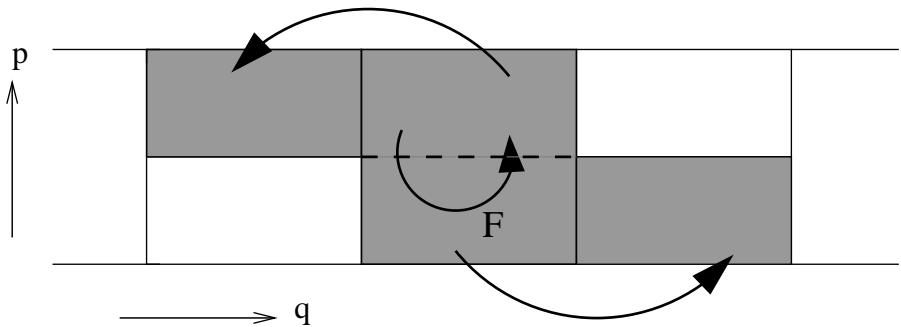}
\caption{The deterministic model of the walk, an intra-cell dynamics $F$ is 
followed by horizontally partitioning each cell and shifting the top part 
left and the bottom part right.
\label{celldynamics}}
\end{figure}

The general class of walks we will consider are deterministic
classical dynamical systems whose quantization is given by
Eq.~(\ref{defineQRW}). The classical phase-space consists of many
cells arranged linearly side by side, with  intra-cell dynamics say
specified by $F$. Each cell as a whole indicates a ``site'', and the
walk is cell-averaged to give a site to site hopping. Inter-cell
dynamics is specified by a binary choice of a cell partition, one
being shifted right and other left. The quantization of $F$ is the
unitary operator $U$ in Eq.~(\ref{defineQRW}), while the choice of
the partitions will determine $P_R$ and $P_L$. The complexity of
the coin toss is replaced by the complexity of the classical
intra-cell transformation $F$. A schematic illustration of this is
shown in Fig. \ref{celldynamics}, where the partitions have been
chosen as equal horizontal splicings. The quantization
of the entire map including the shifts is the quantum random walk
in Eq.~(\ref{defineQRW}). However when we say that, we have made a
separation between inter-cell and intra-cell dynamics while both of
these occur in the same phase-space and therefore this
quantization does not take care of this fact, as noted above.

To illustrate what we mean by the effect of coin dynamics, let us
consider the well-studied case of the Hadamard coin. This
corresponds to a unitary matrix:
\beq U=\f{1}{\sqrt{2}}\left( \begin{array}{cc}1&1\\1 &-1
\end{array} \right),\eeq
in the standard basis which we call $\{|\uparrow \kt,|\downarrow
\kt \}$. The projectors are projectors on the standard basis,
$P_R=|\ua\kt \br \ua| $ and $P_L=|\da \kt \br \da|$. The coin
dynamics is of course extremely simple, it being a rotation, such
that $U^4=I$ (since $M=2$, in fact $U^2=I$). We generalize this
coin by making it higher dimensional and replacing it with the
discrete Fourier transform. Thus we take:
\beq (U)_{\al \be}=\f{1}{\sqrt{M}} \exp \left( 2 \pi i \al \be/M
\right);\;\;\; \al,\be =0,\ldots,M-1.\eeq
We will use Greek letters to denote states from the cell while
Latin letters will refer to the lattice or site states. We note
that the above reduces to the Hadamard coin for $M=2$. Also the
projectors now generalize to $P_L=\sum_{\al=0}^{M/2-1}|\al\kt \br
\al |$ and $P_R=\sum_{\al=M/2}^{M-1}|\al\kt \br \al |$. We think of
the classical coin dynamics as acting on a unit cell $(q,p) \,\in
\,[0,1) \times [0,1)$, while the quantum dynamics is its
quantization on this cell after assuming periodic (or
quasi-periodic) boundary conditions on $q$ and $p$. Thus $U_{\al
\be}$ has the simple interpretation of a quantization of the
canonical transformation $F(q,p) = (1-p,q)$ which is a rigid
anti-clockwise cell rotation by ninety degrees.

\begin{figure}
\includegraphics[height=4in]{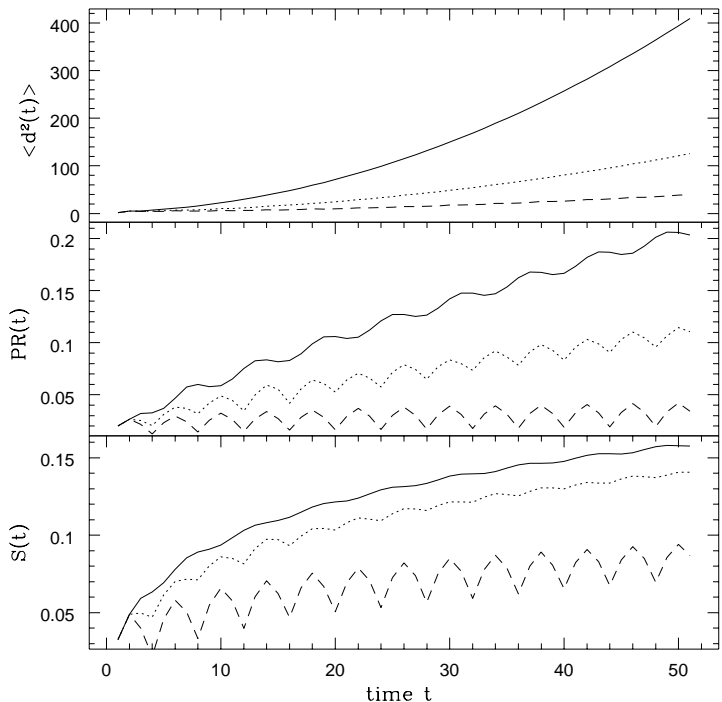}
\caption{The mean square displacement (m.s.d.), participation ratio and the entropy as a function of time for $M=2$ (solid line), $M=10$ (dotted line) and $M=40$ (dashed line) for the Hadamard-Fourier walk. $L=100$ uniformly.
\label{dft1}}
\end{figure}

 Thus replacing the Hadamard coin with the Fourier transform has a
 very simple classical limit, a limit that clearly engenders no
 randomness. This is true even of the walk as a whole, not only for
 the coin dynamics. The simple coin dynamics engenders a simple walk,
 in fact the walker does not go very far. Dividing the individual
 cells into four equal squares, it is easy to see that the walker
 returns exactly to the beginning point after four time steps. Thus
 this limit of the Hadamard walk characterized by the size of the
 Hilbert space $M$ must be such that that $E^4(M \rightarrow
 \infty)=I$, implying a tendency to four-fold spectral degeneracy, and
 a general reluctance to walk.  This is illustrated in
 Fig. \ref{dft1}, before which we introduce some measures. The
 probability of starting at cell or site $0$ and arriving at cell $l$
 after a time $t$ is
\beq p_l(t)=\f{1}{M}\mbox{tr}\left(P_l  E^{-t} P_0
E^{t}\right)\eeq
where $P_i$ is the projection operator on cell $i$. Note that
since the cells are linearly arranged, there exists unambiguous
quantum projectors such as these. The numbers $p_l(t)$ are a
discrete probability measure such that $\sum_{l=0}^{L-1}p_l(t)=1$.
The diffusion in the sites has been usually studied as the second
moment of this distribution, the mean squared displacement (m.s.d.):
\beq <d^2(t)>=\sum_{l=0}^{L-1} p_l(t) d_l^2,\eeq
where $d_l$ is the number of cells from the cell number $0$. This
is however partial as it does not properly distinguish generic
ballistic motions from simple non-random walks such as plain
hopping. Thus the $t^2$ law behind the quantum random walks may
also be a simple non-random dynamics. As a measure of how many
sites are being simultaneously  accessed, we need to know how many
of the $p_l(t)$ at any given time are significant. Thus we may
either use an entropy measure $S(t)$ or a participation ratio measure $PR(t)$:
\beq S(t)=-\sum_{l=0}^{L-1} p_l(t) \log_L(p_l(t)),\;\,
PR(t)=\f{1}{L \sum_{l=0}^{L-1} p_l^2(t)}. \eeq
The measure $PR(t)$ is the fraction of cells that are occupied at
time $t$. We note that for simple hopping while the second moment
is quadratic the entropy is zero and the participation ratio is
$1/L$.

We also have to point out the following generality. The dimension of the
Hilbert space is $ML$, but the uniformity of the coin over the lattice
implies that there is translational invariance and hence the
lattice-momentum will be conserved and this basis will block
diagonalize the unitary operator $E$ into $L$ blocks of size $M$
each. The lattice momentum basis diagonalizes the lattice shift
operator $S$:
\beq S|k\kt =e^{2\pi i k/L}|k\kt.\eeq
The vectors $|k\kt$ are such that $k=0,\ldots,L-1$ and $\br
n|k\kt=\exp(2\pi i nk/L)/\sqrt{L}.$ Then in the lattice-momentum
and a cell basis $\{|\al\kt\}$ that could be any orthonormal set,
the random walk operator $E$ is block diagonal:
\beq \br k, \al|E|k,\be\kt = \left\{ \begin{array}{ll} e^{2 \pi i
k/L} \br \al|U|\be\kt & 0 \le \al \le M/2-1\\e^{-2 \pi i k/L} \br
\al |U|\be \kt & M/2 \le \al \le M-1. \end{array}\right.\eeq
We will choose $\{|\al\kt\}$ to be either position or momentum
states, thereby partitioning the cells either vertically or
horizontally. Thus although the random walk operator $E$  is $LM$
dimensional we can reduce it to $L$, $M$-dimensional operators.

In Fig.~\ref{dft1}, all the three quantities, the m.s.d., the
participation ratio and the entropy, are shown for the
Hadamard-Fourier walk introduced above and we see that as $M$
increases, indeed the walker gets increasingly lethargic and does not
go very far. The oscillatory features seen clearly for $M=10$ and $40$
correspond to period of four and are quantum precursors of the
classically exact period-$4$ behaviour. We note that the partitions
taken here splice the cells equally vertically rather than
horizontally, but this is an inessential feature. Thus the classical
limit of the walk with the Hadamard coin via the Fourier coin leads to
stunted walks and has its origins in the non-randomness of the
classical-coin limit which is a simple rational rotation.  This does
not mean that the Hadamard Walk in itself has an integrable classical
coin limit, as many other coin operators can limit to the Hadamard
matrix for $M=2$, other than the Fourier transform. For instance the
quantum baker map itself can be a finite $M$ generalization of the
Hadamard coin, as pointed out in \cite{DD2}. Thus in extreme quantum
cases, the notions of integrability and nonintegrability would lose
significance. We now turn to cases where the large coin space limit
is non-trivial, in fact nonintegrable.

\section{Nonintegrable coins}

We now wish to consider classical dynamics of coins $F(g)$ that
depend on a parameter $g$ which can drive it from integrability to
chaos. The degree of randomness in the walk will then be
controlled by the parameter $g$. In the earlier section we had
considered a simple coin which was an integrable example of a
rotation. There exist a wide class of models we can choose from,
including the so called standard map \cite{LL}, but we will pick the Harper
map, we expect much of what we describe to be independent of such
choices.

First, however we briefly mention the exactly solvable case of the
multi-baker which has been mentioned earlier. In this case the cell
dynamics is the baker transformation:
\beq F(q,\,p)=\left\{\begin{array}{ll} (2q,\,p/2)&0\le q
<1/2\\(2q-1,\,(p+1)/2) & 1/2\le q <1.\end{array}\right.\eeq
 In the infinite chain of such bakers, connected by shifts as
 described above, if one cell (say cell $1$) is ``filled'' uniformly
 (more formally, the initial density is the characteristic function
 over this cell) the amount of overlap with the other cells will be
 same as the probability of an unbiased random walker being at that
 cell, if she were to start from cell $1$. Thus in a very real sense
 the quantization of such maps are quantum random walks. These have
 been studied to some extent \cite{DanielDorfman} but we wish to
 examine coins that show a range of dynamical behaviour, through a
 controllable parameter.  It may be noted that to the best of our
 knowledge even classical studies on such systems (``multi-standard'',
 ``multi-Harper'') has not been done and represent rich models from
 many perspectives in physics.

\begin{figure}
\includegraphics[height=4in,width=4in ]{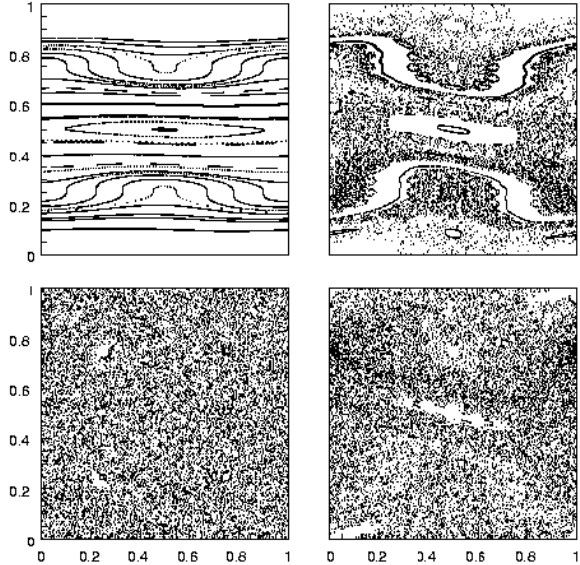}
\caption{The phase space $(q,p)$ of the classical map for $\tau=1$ and
$g=0.01$, $0.05$, $0.1$, and $1$ clockwise from top left.
\label{harpmap}}
\end{figure}

The Harper map we use subsequently is given by the following
transformation:
 \beqa
q_{t+1}&=&q_t- \tau \sin(2\pi p_t)\nonumber\\ p_{t+1}&=&p_t + \tau
g \sin(2 \pi q_{t+1}), \eeqa where $t$ is integer time, and
$(q,p)$ is on a unit torus, so modulo one operation is assumed.
This is a two-parameter area preserving transformation, and has
been studied by many authors for various purposes \cite{Harper}. 
This can be derived from the equations of motion of the 
time-dependent Hamiltonian
\beq
H=\cos(2\pi p)+g \cos(2\pi q) \sum_{n=-\infty}^{\infty} \delta(2 \pi t/\tau-n),
\eeq
as the map connecting the states just after consecutive kicks. One of
the possibilities that seems studied is that for which $g=1$ and the
parameter $\tau$ varies. As $\tau/2\pi$ is the time between kicks this
alters time-scales in the problem, and therefore we choose to fix
$\tau=1$ and consider the Harper map as a function of $g$. Since this
phase-space does not seem familiar, we show in fig.~\ref{harpmap} some
cases, which illustrates a transition to chaos. The routes as $g$
increases seems quite non-trivial as regions of regularity can be
created by bifurcations, but by and large the system is almost
completely chaotic beyond $g=1$. It may be noted that although we are
only displaying the phase-space of $F$, a single cell dynamics, this
reflects the phase-space of the complete walker as this will be
multiple copies of the same. The somewhat unusual structures visible
at $g=0.01$ is because at $\tau=1$ the fixed point(s) at
$(q=1/2,p=1/4,3/4)$ is born, and hence this is a marginal
situation. We may say in the context of this work that as $g$
increases, the coin becomes increasingly random and the walk it
generates will reflect this in ways we want to study.

As is standard the Floquet operator (quantum map) is the quantum
propagator: \beq U(g)= \exp(-i \tau g \cos(2 \pi \hat{q})/h)
\exp(-i \tau \cos(2 \pi \hat{p})/h). \eeq With $h=1/M$, we get the
quantum version in the basis spanned by the momentum states
$|\al\kt$:
\beqa &&\br \al |U(g) |\be \kt = \exp\left(-i \tau M \cos(2\pi
(\be +\phi)/M)\right)\nonumber \\ &&\f{1}{M}\sum_{\ga=0}^{M-1}
\exp\left(-i \tau g M \cos(2\pi (\ga +\phi)/M)\right)
\exp\left(2\pi i (\ga+\phi)(\be -\al)/M\right).\eeqa
We have used here the transformation function:
\beq \br \ga|\al\kt = \f{1}{\sqrt{M}} \exp\left(2 \pi i
(\ga+\phi)(\al+\phi)/M\right),\eeq
where $|\ga \kt$ is a coin position eigenket and $|\al\kt$ is a
coin momentum state. The angle $\phi$ refers to boundary
conditions on the $M$ coin states and here we have assumed that
both position and momentum states acquire a phase of $\exp(2\pi i
\phi)$ for a translation of a cell (or $M$ states). We could have
assumed two different phases here, but for simplicity we introduce
equal phases in both position and momentum states, this phase can
be used to break time-reversal (TR) symmetry for the coin
dynamics, the effect of which will be felt on the random walker.
In the context of the dynamical model we may think of the phase as
an Aharanov-Bohm phase that has no effect on the classical
dynamics.

We now make use of this unitary coin in the random walk Eq.
~(\ref{defineQRW}) and find
\beq p_{l}(t)=\f{1}{M}\sum_{\al=0}^{M-1}\sum_{\be=0}^{M-1} |\br
0,\al|E^t|l,\be\kt|^2.\eeq
We then find the entropy, PR, and m.s.d. on the sites as a function of
the chaos parameter $g$, coin space dimensionality $M$, and the TR
symmetry breaking phase $\phi$. For the rest of the paper we set
$\tau=1$ fixing a definite time scale.

\begin{figure}
\includegraphics[height=4in ]{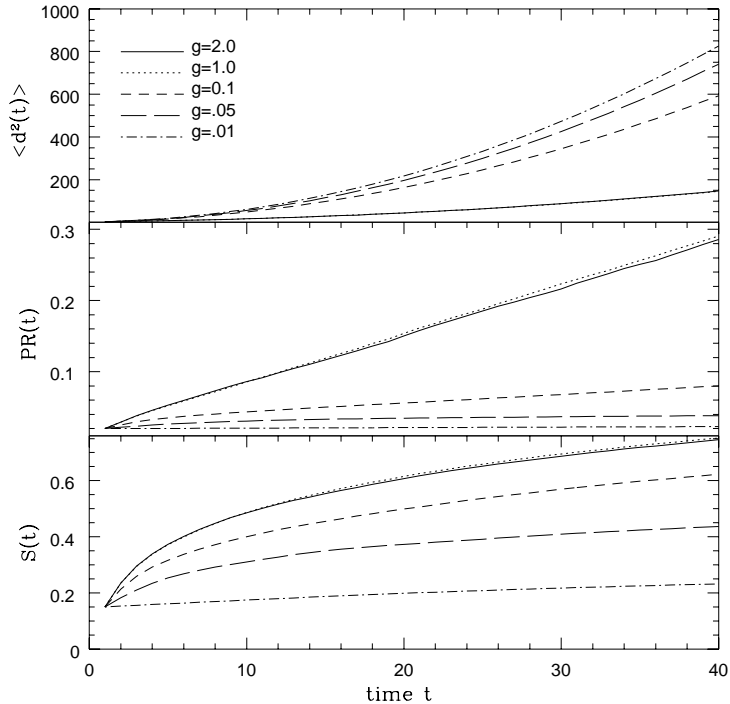}
\caption{The m.s.d., the participation ratio and the entropy as function of time for the multi-Harper walk for various values of the chaos parameter $g$. Uniformly $M=20$ and $L=100$.
\label{fig4.ps}}
\end{figure}

In Fig.~\ref{fig4.ps} the various quantities are plotted as a
function of time for five representative values of the chaos
parameter $g$. We have only shown the first $40$ time steps so
that effects of the finite size of the lattice (uniformly $L=100$)
is minimized, and is practically independent of this. It is seen
that while the m.s.d. is large for the near integrable coin
dynamics (small $g$), it is the opposite when considering the PR
and entropy of the walk. Both of these quantities are large and
are practically the same once chaos has been achieved. This is
shown as the near coalescence of the $g=1$ and $g=2$ curves. Thus
this figure illustrates the role of deterministic chaos on the
quantum walk. We also note the near linear increase in the PR with
time in the chaotic cases and the intermediate behavior for mixed
coin dynamics. Obviously more detailed study of these features is
desirable, but is not pursued here.

\begin{figure}
\includegraphics[height=4in ]{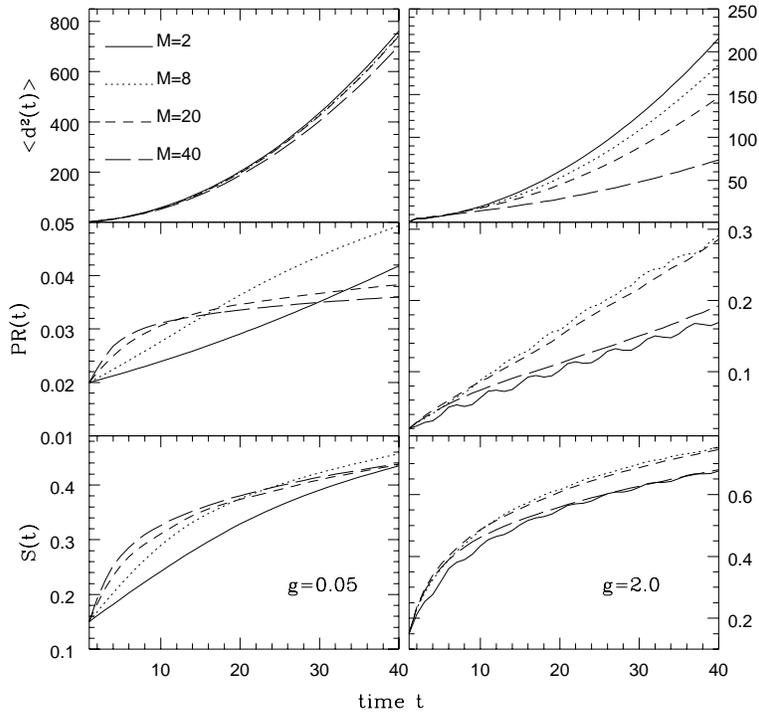}
\caption{The m.s.d., the participation ratio and the entropy as function of time for the multi-Harper walk for various values of the coin dimensionality $M$.
Shown are one near-integrable (left) and one chaotic (right) cases. 
The size of the lattice is $L=100$.
\label{dim}}
\end{figure}

Variation of the coin dimensionality $M$ can have a strong impact on
the quantum walk and we turn attention in Fig.~\ref{dim} to this. For
a near-integrable case it is seen that the coin-dimensionality does
not have a significant impact on the m.s.d. which increases
quadratically, however significant deviations are seen for the PR and
the entropy. No simple rules are apparently operative, except that for
large enough $M$ there is a convergence. For a chaotic case, the
m.s.d.  is clearly larger for $M=2$ in comparison with larger coins,
however PR and entropy wise the larger dimensionality is preferred,
although the relationship is not monotonic, for the cases shown $M=8$
seems to produce the most PR and entropy for the times calculated herein.

\begin{figure}
\includegraphics[height=4in ]{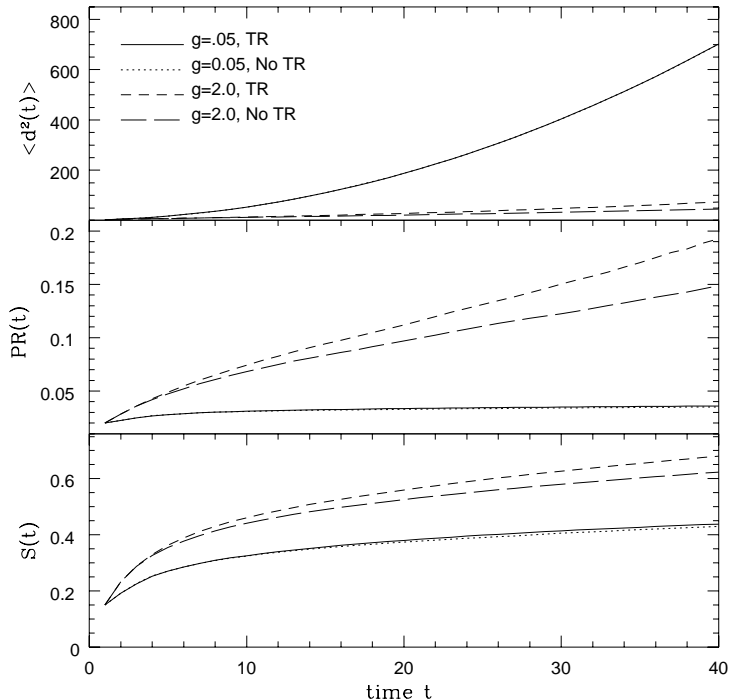}
\caption{The m.s.d., the participation ratio and the entropy as function of 
time for the multi-Harper walk with $(\phi=0)$ and without
time-reversal symmetry $(\phi=0.2)$, for a case that is
near-integrable ($g=0.05$) and one that is chaotic ($g=2$). In all
cases $M=40$ and $L=100$.
\label{fig6.ps}}
\end{figure}

In Fig.~\ref{fig6.ps} is shown the effect of TR symmetry breaking
coins. We break the symmetry quite simply by taking a non-zero phase
$\phi$, in this case $0.2$. For the nearly integrable/mixed case when
$g=0.05$, the TR symmetry breaking has practically no effect while a
large effect is seen when there is classical chaos. In the case when
the coin dynamics has lost its TR symmetry the walk tends to be slower
and produces less entropy.  This is because of the additional
destructive interference from time reversed paths. The sensitivity to
TR symmetry is a general feature of quantum chaos
\cite{Haake}. For instance in the context of information physics and
chaos it was recently shown that TR symmetry has a crucial role in
determining the distribution of entanglement present in a certain
class of many-particle states
\cite{ArulSubbu}. Thus we expect that TR symmetry breaking leads to a
decreased diffusion. It must be noted that this is a purely quantum
phenomenon, the classical dynamics is completely unaltered by the
phases, in fact for initial times the TR symmetric and non-TR
symmetric cases go hand-in-hand, a period that agrees with classical
laws.

\section{Discussion}

In this paper we have begun the study of a class of deterministic
quantum random walkers that generalizes the classical multi-baker maps
of Tasaki and Gaspard. These maps show a mixture of regular and
chaotic dynamics, and their quantization is in the canonical
form of simple random walks. Our main query was what is random in a
quantum random walk? In the classical context clearly the source of
randomness is the coin-toss and not any process of measurement.  Thus
we believe that any randomness in quantum random walks must come from
using quantum chaotic coins, meaning simply quantized classically
chaotic coins. The classical limit in these cases corresponds to the
classical limit of the coins, in which case the classical diffusion
laws will be obtained. In fact the classical dynamics of such systems
as introduced here, namely the multi-Harper maps is itself not studied
and represents a potentially rich source of transport models. We have
first shown how the generalization of the much studied Hadamard walk
is in fact a simple Fourier walk that tends to ``go nowhere'', due to
its eventual, classical, period-4 behaviour.

We have introduced two natural quantities in studying the quantum
walks, the site entropy and the site participation ratio, which
measure how well the probability distribution is sampling the sites,
rather than only the m.s.d. which tells us how far the walker had
gone. In fact we have found that while the m.s.d. is large for nearly
regular coins, the entropy and PR are small, indicating the simple
nature of the quantum diffusion. On the other hand for chaotic coins
the PR and entropy are large and continue to be produced, with the PR
increasing linearly in time.  We have also shown that TR symmetry
breaking suppresses the quantum walk when the coin is chaotic, the
most interesting case. We have only aimed at introducing the models
and showing primarily a few central numerical results, it is believed
that these are of sufficient interest to warrant further study.

\begin{acknowledgments}
This work was not funded by agencies either involved in the
construction or alleged destruction of weapons of mass destruction. It
has solely been financed by the tax payers of India via the agency of the 
Department of Space, Govt. of India.
\end{acknowledgments}


\begin{thebibliography}{99}


\bibitem{kempe} J. Kempe, quant-ph/0303081.

\bibitem{Nayak} A. Nayak, A. Vishwanath, quant-ph/0010117, and DIMACS Technical Report 2000-43.

\bibitem{Ornstien} D. Ornstein, Science {\bf 243}, 182 (1989).

\bibitem{LL} A. J. Lichtenberg, and M. A. Lieberman, {\it Regular and Chaotic 
Dynamics}, 2nd ed., Springer-Verlag (New York, 1992).  

\bibitem{BalVoros} N. L. Balazs, and A. Voros, Ann. Phys. (N. Y.) {\bf 190}
(1989) 1; M. Saraceno, Ann. Phys. (N.Y.) {\bf 199}, 37, (1990).

\bibitem{DanielDorfman} D. K. Wojcik, J. R. Dorfman, Phys. Rev. E {\bf 66}, 036110 (2002); quant-ph/0209036.

\bibitem{DD2} D. K. Wojcik, J. R. Dorfman, quant-ph/0212036. 

\bibitem{Gasp} P. Gaspard, J. Stat. Phys. {\bf 68}, 673 (1992).

\bibitem{GT} S. Tasaki, P. Gaspard, J. Stat. Phys. {\bf 101}, 125 (1995).

\bibitem{ArulBal} A. Lakshminarayan, N. L. Balazs, J. Stat. Phys. {\bf 77}, 311 (1994).

\bibitem{Harper} R. Artuso, G. Casati,  F. Borgonovi, L. Rebuzzini, Int. J. Mo
d. Phys. B {\bf 8} 207 (1994); P. Leboeuf, J. Kurchan, M. Feingold and
D. P. Arovas, Phys . Rev. Lett. {\bf 65} 3076 (1990); R. Lima and
D. Shepelyansky, Phys. Rev. Lett.  {\bf 67}, 1377 (1991).

\bibitem{ArulSubbu} A. Lakshminarayan, V. Subrahmanyam, Phys. Rev. A. 
To Appear (May 2003). quant-ph/0212049.
 
\bibitem{Haake}  F. Haake, {\it Quantum Signature of Chaos} 2nd Ed.
Springer-Verlag (Berlin, 2001).


\end{thebibliography}
\end{document}